\documentstyle[12pt,epsfig]{article}
\begin{document}

\begin{flushright}
{\bf hep-ph/0007136} \\
LMU-00-08
\end{flushright}

\vspace{0.4cm}

\begin{center}
{\large\bf Final-state Rescattering Effects on 
$B_d\rightarrow \pi\pi$ Decays and CP Violation}
\end{center}

\vspace{.5cm}
\begin{center}
{\bf Zhi-zhong Xing} \footnote{
E-mail: Xing$@$hep.physik.uni-muenchen.de }\\
{\small\sl Sektion Physik, Universit${\sl\ddot a}$t M${\sl\ddot u}$nchen, 
Theresienstrasse 37A, 80333 M${\sl\ddot u}$nchen, Germany}
\end{center}

\vspace{3.5cm}

\begin{abstract}
The significance of final-state interactions
in $B_d\rightarrow \pi\pi$ decays is phenomenologically demonstrated 
by taking elastic $\pi\pi\rightleftharpoons \pi\pi$ and
inelastic $\pi\pi\rightleftharpoons D\bar{D}$ rescattering effects
into consideration. We find that the present experimental data on
$B^0_d\rightarrow \pi^+\pi^-$ can well be understood in this
approach without fine-tuning of the input parameters, and large 
CP-violating asymmetries are expected to manifest themselves in
such charmless rare processes. 
\end{abstract}

\newpage

Recently the branching ratio of $B^0_d\rightarrow \pi^+\pi^-$ has
been measured independently by CLEO, BABAR and BELLE Collaborations:
\begin{equation}
{\cal B}(B^0_d\rightarrow \pi^+\pi^-) \; = \; 
\left \{ \matrix{
\left (4.3^{+1.6}_{-1.4} \pm 0.5 \right ) \times 10^{-6} 
~~~({\rm CLEO \cite{CLEO}}) \; , \cr
~ \left (6.3^{+3.9}_{-3.5} \pm 1.6 \right ) \times 10^{-6} 
~~~({\rm BELLE \cite{BELLE}}) \; , \cr
\left (9.3^{+2.6+1.2}_{-2.3-1.4} \right ) \times 10^{-6} 
~~~({\rm BABAR \cite{BABAR}}) \; . \cr
} \right .
%               (1)
\end{equation}
Theoretical predictions for ${\cal B}(B^0_d\rightarrow \pi^+\pi^-)$,
as those given in Refs. \cite{Beneke,Deshpande,Du,Beneke2}
based on the QCD-improved factorization, are in good agreement
with the BABAR data but difficult to coincide with the CLEO data.
It has to be seen, in the near future, how three measurements will 
reach full consistency and whether the final experimental result
can well be understood in the factorization approach without
fine-tuning of the input parameters.
 
If the present CLEO data are taken seriously, it seems
necessary to modify the theoretical 
prediction for ${\cal B}(B^0_d\rightarrow \pi^+\pi^-)$.
To do so one naturally speculates 
that final-state interactions in $B\rightarrow \pi\pi$ decays
might be significant and ought not to be ignored in the
factorization approach. 
In Refs. \cite{Hou,Wu} the elastic
$\pi\pi \rightleftharpoons \pi\pi$ rescattering effects on the branching
ratios and CP asymmetries of $B\rightarrow \pi\pi$ transitions have been
demonstrated to be important. It is also likely that
such rare nonleptonic processes are deeply involved in
inelastic final-state interactions \cite{Wolfenstein,Sehgal}. However, it
remains a big challenge today to handle the inelastic rescattering 
effects on $B$ decays in a systematic and quantitative way \cite{Peccei}.

In this paper we attempt to follow a purely phenomenological approach
to demonstrate the influence of inelastic final-state interactions on 
$B_d\rightarrow \pi\pi$ decays and CP violation. The essential
argument is that there may {\it a priori} exist a two-step process 
$B_d \rightarrow D\bar{D}\Longrightarrow \pi\pi$, arising from inelastic
$\pi\pi \rightleftharpoons D\bar{D}$ rescattering, in addition to the
direct decay mode $B_d \rightarrow \pi\pi$. 
We find that both elastic and inelastic
rescattering effects are possible to modify the predictions based on the
QCD-improved factorization. More precise measurements of 
$B\rightarrow \pi\pi$ decays are expected to clarify whether 
final-state interactions in them are really significant or not.

It is well known that the final state $\pi^+\pi^-$ or $\pi^0\pi^0$
of $B_d$ decays consists of both $I=0$ and $I=2$ isospin configurations, and 
$D^+D^-$ or $D^0\bar{D}^0$ consists of both $I=0$ and $I=1$
isospin configurations. Under inelastic rescattering the $I=0$
configuration of $\pi\pi$ can mix with that of $D\bar{D}$, leading
to a two-step decay mode $B_d \rightarrow D\bar{D} \Longrightarrow
\pi\pi$. The final states $\pi^\pm \pi^0$ of $B^{\pm}_u$ decays,
which only have the $I=2$ isospin configuration, 
cannot mix with $D\bar{D}$. 
But $\pi^\pm \pi^0$ are possible to mix with the final states like
$\rho^\pm \rho^0$, and $\pi^+\pi^-$ (or $\pi^0\pi^0$) could also mix with 
the final states such as $K^+K^-$ and $K^0\bar{K}^0$ \cite{Sehgal}.
For simplicity, we assume that the inelastic final-state interactions
of $B_d \rightarrow \pi\pi$ decays are dominated by the $I=0$ channel
mixing via $\pi\pi \rightleftharpoons D\bar{D}$ scattering,
leaving the $I=2$ state of $\pi\pi$ unmixed with others. 
In the assumption made above and in the neglect of small
electroweak penguin contributions, the amplitudes of
$B^0_d \rightarrow \pi^+\pi^-$, $B^0_d \rightarrow \pi^0\pi^0$,
and $B^+_u \rightarrow \pi^+\pi^0$ decay modes can be 
written as \cite{Xing95} 
%%%%%%%%%%%%%%%%%%%%%%
\footnote{Note that the sign of $A^{\pi\pi}_0$ 
is here taken to be different from that
in Ref. \cite{Xing95}. The present choice will prove convenient
when the factorization approximation is applied to the 
isospin amplitudes.}:
%%%%%%%%%%%%%%%%%%%%%%
\begin{eqnarray}
A(B^0_d \rightarrow \pi^+\pi^-) & = & \sqrt{2} S^{\pi\pi}_2 A^{\pi\pi}_2 + 
\sqrt{2} \left ( S^{\pi\pi}_0 A^{\pi\pi}_0 + 
S^{\pi D}_0 A^{D\bar D}_0 \right ) \; , 
\nonumber \\
A(B^0_d \rightarrow \pi^0\pi^0) ~ & = & 
2 S^{\pi\pi}_2 A^{\pi\pi}_2 - \left ( S^{\pi\pi}_0 A^{\pi\pi}_0 
+ S^{\pi D}_0 A^{D\bar D}_0 \right ) \; ,
\nonumber \\
A(B^+_u \rightarrow \pi^+\pi^0) & = & 3 A^{\pi\pi}_2 \; ,
%               (2)
\end{eqnarray}
where $A^{\pi\pi}_{0,2}$ denote the ``bare'' isospin amplitudes of 
$B \rightarrow \pi\pi$, $A^{D\bar D}_0$ stands for the $I=0$ isospin
amplitude of $B_d\rightarrow D\bar{D}$ \cite{Xing99}, 
$S^{\pi\pi}_{0,2}$ and
$S^{\pi D}_0$ are the inelastic-rescattering matrix elements connecting 
the unitarized isospin amplitudes to the bare ones \cite{Sehgal}. 
Obviously $S^{\pi D}_0 =0$ and $S^{\pi\pi}_2 = S^{\pi\pi}_0 =1$ held, 
if there were no mixture between the $I=0$ states of $\pi\pi$ and $D\bar{D}$. 

The bare isospin amplitudes $A^{\pi\pi}_{0,2}$ and $A^{D\bar D}_0$ can be
calculated with the help of the effective weak Hamiltonian \cite{Buras}
and the QCD-improved factorization \cite{Beneke2}. After a straightforward
calculation, we obtain
\begin{eqnarray}
A^{\pi\pi}_0 & = & \frac{G_{\rm F}}{6} \left [ V^*_{ub} V_{ud}
\left ( 2 a_1 - a_2 + 3 a^u_4 + 3 a^u_6 \xi^{~}_\pi \right )
+ V^*_{cb} V_{cd} \left ( 3 a^c_4 + 3 a^c_6 \xi^{~}_\pi \right )
\right ] T_{\pi} e^{i\delta_0} \; ,
\nonumber \\
A^{\pi\pi}_2 & = & \frac{G_{\rm F}}{6} V^*_{ub} V_{ud} (a_1 + a_2)
T_{\pi} e^{i\delta_2} \; ,
\nonumber \\
A^{D\bar D}_0 & = & \frac{G_{\rm F}}{\sqrt{2}} \left [ V^*_{ub}V_{ud}
\left (a^u_4 + a^u_6 \xi^{~}_D \right ) + V^*_{cb} V_{cd}
\left ( a_1 + a^c_4 + a^c_6 \xi^{~}_D \right ) \right ] T_{D} e^{i\delta_0} \; ,
%             (3)
\end{eqnarray}
in which $V_{ub}$, $V_{ud}$, $V_{cb}$, and $V_{cd}$ are the quark mixing matrix
elements; $a_1$, $a_2$, $a^{u,c}_4$, and $a^{u,c}_6$ are the QCD coefficients
independent of the renormalization scheme \cite{Beneke};
$\xi^{~}_\pi$ and $\xi^{~}_D$ are the factorization parameters arising from
the transformation of ${\rm (V-A)(V+A)}$ currents into 
${\rm (V-A)(V-A)}$ ones for the penguin operators $Q_5$ 
and $Q_6$ \cite{Buras}; $\delta_0$ and $\delta_2$ are strong (isospin) phases,
$T_{\pi}$ and $T_{D}$ denote the factorized hadronic matrix
elements of $B\rightarrow \pi\pi$ and $B\rightarrow D\bar{D}$ decays
respectively. Under isospin symmetry $\xi^{~}_\pi$ and $\xi^{~}_D$ read
\begin{eqnarray}
\xi^{~}_\pi & = & \frac{2m^2_\pi}{(m_b - m_u) ( m_u + m_d)} \; ,
\nonumber \\ 
\xi^{~}_D & = & \frac{2m^2_D}{(m_b - m_c) ( m_c + m_d)} \; .
%             (4)
\end{eqnarray}
In terms of the relevant decay constants and form factors, one
gets
\begin{eqnarray}
T_{\pi} & = & i f_\pi F^{B\pi}_0 (m^2_\pi) 
\left ( m^2_B - m^2_\pi \right ) \; ,
\nonumber \\
T_{D} & = & i f_D F^{BD}_0 (m^2_D) \left (m^2_B - m^2_D \right ) \; .
%             (5)
\end{eqnarray}
It is worth remarking that the isospin amplitudes $A^{\pi\pi}_{0,2}$
and $A^{D\bar D}_0$ have been calculated separately in the factorization
approximation, hence the contribution of $A^{D \bar D}_0$ to 
$A(B_d\rightarrow \pi\pi)$
need in principle be normalized. Such a treatment is however 
unnecessary in the approach under discussion, because the 
normalization factor of $A^{D\bar D}_0$ can always be absorbed into 
the unknown parameter $S^{\pi D}_0$ in Eq. (2). This point will be seen 
more clearly later on.

Following Eq. (2) one may write down the similar isospin relations 
for the amplitudes of $\bar{B}^0_d\rightarrow \pi^+\pi^-$,
$\bar{B}^0_d\rightarrow \pi^0\pi^0$, and $B^-_u\rightarrow \pi^-\pi^0$
decay modes, whose bare isospin amplitudes $\bar{A}^{\pi\pi}_{0,2}$
and $\bar{A}^{D\bar D}_0$ can directly be read off from 
$A^{\pi\pi}_{0,2}$ and $A^{D\bar D}_0$ in Eq. (3) with the 
replacements $V^*_{ub}V_{ud} \Longrightarrow V_{ub}V^*_{ud}$ and
$V^*_{cb}V_{cd} \Longrightarrow V_{cb}V^*_{cd}$. Of course
$|\bar{A}^{\pi\pi}_2/A^{\pi\pi}_2| = 1$ holds in the approximation of
neglecting the electroweak penguin effect. 

Let us give a brief retrospection of the conventional calculations of
$B \rightarrow \pi\pi$ decays, in which final-state interactions
are considered only at the quark level \cite{Beneke,Deshpande,Du,Beneke2}.
Taking $\delta_0 = \delta_2 =0$, $S^{\pi D}_0 = 0$, and
$S^{\pi\pi}_0 = S^{\pi\pi}_2 =1$ (i.e., neglecting both elastic 
$\pi\pi\rightleftharpoons \pi\pi$ and inelastic 
$\pi\pi \rightleftharpoons D \bar{D}$ rescattering effects),
we arrive from Eqs. (2) and (3) at
%%%%%%%%%%%%%%%%%%%%%%%%%%%%%%%%%%
\footnote{Note that there is a factor of 1/2 for the phase space of 
$B^0_d\rightarrow \pi^0\pi^0$ due to identical final-state particles.
This factor has already been absorbed into the transition 
amplitude $A(B^0_d\rightarrow \pi^0\pi^0)$ in Eqs. (2) and (6).}
%%%%%%%%%%%%%%%%%%%%%%%%%%%%%%%%%%
\begin{eqnarray}
A(B^0_d\rightarrow \pi^+\pi^-)_{\bf 0} & = & \frac{G_{\rm F}}{\sqrt{2}}
\left [ V^*_{ub} V_{ud} \left ( a_1 + a^u_4 + a^u_6 \xi^{~}_\pi \right )
+ V^*_{cb} V_{cd} \left ( a^c_4 + a^c_6 \xi^{~}_\pi \right ) \right ] T_\pi \; ,
\nonumber \\
A(B^0_d\rightarrow \pi^0\pi^0)_{\bf 0} ~ & = & \frac{G_{\rm F}}{2} 
\left [ V^*_{ub} V_{ud} \left ( a_2 - a^u_4 - a^u_6 \xi^{~}_\pi \right )
- V^*_{cb} V_{cd} \left ( a^c_4 + a^c_6 \xi^{~}_\pi \right ) \right ] T_\pi \; ,
\nonumber \\
A(B^+_u\rightarrow \pi^+\pi^0)_{\bf 0} & = & \frac{G_{\rm F}}{2} 
V^*_{ub} V_{ud} (a_1 + a_2) T_\pi \; ,
%             (6)
\end{eqnarray}
where the subscript ``{\bf 0}'' denotes the absence of final-state 
interactions at the hadron level.
Such a result has been presented in Refs. \cite{Beneke,Deshpande}. The branching
ratios of $B\rightarrow \pi\pi$ decays can then be computed,
under isospin symmetry, by using the formula
\begin{equation}
{\cal B}(B\rightarrow \pi\pi)_{\bf 0} \; =\; 
\frac{\tau^{~}_B \sqrt{m^2_B - 4 m^2_\pi}}
{16\pi m^2_B} \left | A(B\rightarrow \pi\pi)_{\bf 0} \right |^2 \; ,
%            (7)
\end{equation}
where $\tau^{~}_B$ denotes the $B$-meson lifetime. To obtain the quantitative
results for ${\cal B}(B\rightarrow \pi\pi)_{\bf 0}$, we input the following
typical values of the quark mixing parameters:
$|V_{ud}| = 0.9735$, $|V_{cd}| = 0.224$,
$|V_{cb}| = 0.0402$, $|V_{ub}|/|V_{cb}| = 0.090$, and
$\gamma \equiv \arg(-V^*_{ub}V_{ud}V_{cb}V^*_{cd}) = 65^\circ$ \cite{PDG}.
Furthermore $\tau^{~}_B = 1.6$ ps, $f_\pi = 130.7$ MeV, and 
$F^{B\pi}_0 (m^2_\pi) \approx F^{B\pi}_0(0) = 0.3$ are used \cite{PDG,Cheng}.
Following Beneke {\it et al.} in Ref. \cite{Beneke}, 
we adopt $a_1 = 1.038 + 0.018i$, $a_2 = 0.082 - 0.080i$, $a^u_4 = -0.029 -0.015i$,
and $a^c_4 = -0.034 - 0.008i$ at the scale $\mu = m^{~}_b$, and neglect
the formally power-suppressed QCD coefficients $a^{u,c}_6$ in the
heavy quark limit. The predictions for the branching ratios of
$B\rightarrow \pi\pi$ turn out to be
%%%%%%%%%%%%%%%%%%%%%%%%%%
\footnote{The upper bound on the branching ratio of
$B^0_d\rightarrow \pi^0\pi^0$ has been reported by the CLEO
Collaboration \cite{CLEO}: 
${\cal B}(B^0_d\rightarrow \pi^0\pi^0) < 5.7 \times 10^{-6}$; and
the upper bounds on the branching ratio of $B^\pm_u\rightarrow \pi^\pm\pi^0$
have been reported by both CLEO and BELLE Collaborations:
${\cal B}(B^\pm_u\rightarrow \pi^\pm\pi^0) < 1.27 \times 10^{-5}$
(CLEO \cite{CLEO}) and 
${\cal B}(B^\pm_u\rightarrow \pi^\pm\pi^0) < 1.01 \times 10^{-5}$
(BELLE \cite{BELLE}).}
%%%%%%%%%%%%%%%%%%%%%%%%%%
\begin{eqnarray}
{\cal B}(B^0_d\rightarrow \pi^+\pi^-)_{\bf 0} & \approx & 
1.0 \times 10^{-5} \; ,
\nonumber \\
{\cal B}(B^0_d\rightarrow \pi^0\pi^0)_{\bf 0} ~ & \approx & 
1.4 \times 10^{-8} \; ,
\nonumber \\
{\cal B}(B^+_u\rightarrow \pi^+\pi^0)_{\bf 0} & \approx & 
5.8 \times 10^{-6} \; ;
%            (8)
\end{eqnarray}
and
\begin{eqnarray}
{\cal B}(\bar{B}^0_d\rightarrow \pi^+\pi^-)_{\bf 0} & \approx & 
9.8 \times 10^{-6} \; ,
\nonumber \\
{\cal B}(\bar{B}^0_d\rightarrow \pi^0\pi^0)_{\bf 0} ~ & \approx & 
1.5 \times 10^{-7} \; ,
\nonumber \\
{\cal B}(B^-_u\rightarrow \pi^-\pi^0)_{\bf 0} & \approx & 
5.8 \times 10^{-6} \; .
%            (9)
\end{eqnarray}
First of all we observe that ${\cal B}(B^0_d\rightarrow \pi^+\pi^-)_{\bf 0}$
deviates almost a factor of 2 from the CLEO data given in Eq. (1), but
it is in good agreement with the preliminary BABAR data. At present it 
remains too early to claim any discrepancy between the theoretical
prediction and the experimental measurements. 
Secondly, the branching ratios of $B^0_d\rightarrow \pi^0\pi^0$ and 
$\bar{B}^0_d\rightarrow \pi^0\pi^0$ are rather
sensitive to the values of the QCD coefficients and the weak phase $\gamma$.
The implication of ${\cal B}(B^0_d\rightarrow \pi^0\pi^0)_{\bf 0}/
{\cal B}(\bar{B}^0_d\rightarrow \pi^0\pi^0)_{\bf 0} \approx 0.1$ is
quite clear: the direct CP-violating asymmetry between these two decay modes
might be of ${\cal O}(1)$. In addition,
$B^{\pm}_u \rightarrow \pi^{\pm}\pi^0$ transitions can be calculated
in a relatively reliable way at the tree level, thus a measurement of
their branching ratios will test the validity of the QCD-factorization
approximation. Actually the preliminary CLEO data 
${\cal B}(B^{\pm}_u \rightarrow \pi^{\pm}\pi^0) = 
(5.4^{+2.1}_{-2.0}\pm 1.5) \times 10^{-6}$ \cite{CLEO}, 
though not yet formally announced, do agree very well with the 
theoretical prediction.
%%%%%%%%%%%%%%%%%%%% Fig. 1 %%%%%%%%%%%%%%%%
\begin{figure}[t]
\epsfig{file=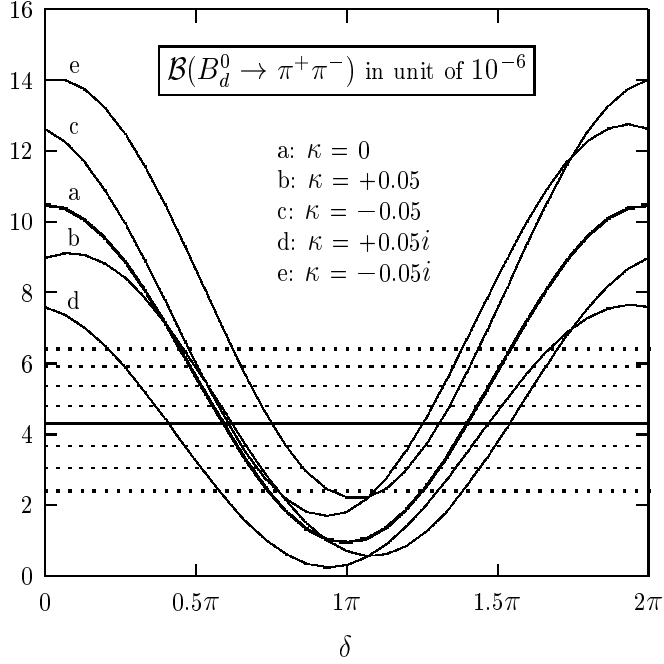,bbllx=1cm,bblly=10cm,bburx=18cm,bbury=27.5cm,%
width=15.5cm,height=17cm,angle=0,clip=}
\vspace{-7.8cm}
\caption{The branching ratio of $B^0_d\rightarrow \pi^+\pi^-$
predicted in the factorization approximation with
different input values of the inelastic rescattering parameter
$\kappa$ and the isospin phase shift $\delta$. The
dashed region is favored by the CLEO data \cite{CLEO}.}
\end{figure}
%%%%%%%%%%%%%%%%%%%%%%%%%%%%%%%%%%%%%%%%%%%%

Now we recalculate the branching ratios of $B_d\rightarrow \pi^+\pi^-$
and $B_d\rightarrow \pi^0\pi^0$ decays by taking the final-state
rescattering effects into account. We assume that the inelastic rescattering
matrix $S$ is approximately diagonal 
(i.e., $S^{\pi\pi}_2 \approx S^{\pi\pi}_0 \approx 1$) and its off-diagonal
element $S^{\pi D}_0$ is only of ${\cal O}(10^{-2})$ in magnitude.
In this reasonable assumption, which should not be far away from reality, one can 
simplify Eq. (2) and arrive at
\begin{eqnarray}
A(B^0_d\rightarrow \pi^+\pi^-) & \approx & \frac{G_{\rm F}}{\sqrt{2}}
\left \{ V^*_{ub} V_{ud} \left [ \frac{1}{3} (a_1 + a_2) e^{i\delta}
+ \frac{1}{3} (2a_1 - a_2) + \left (a^u_4 + a^u_6 \xi^{~}_\pi \right ) \right ]
\right . \nonumber \\
&  & \left . ~~~~ + V^*_{cb} V_{cd} \left [ \sqrt{2} ~ a_1 \kappa +
\left ( a^c_4 + a^c_6 \xi^{~}_\pi \right ) \right ] \right \} 
T_\pi e^{i\delta_0} \; ,
\nonumber \\ \nonumber \\
A(B^0_d\rightarrow \pi^0\pi^0) ~ & \approx & \frac{G_{\rm F}}{2} 
\left \{ V^*_{ub} V_{ud} \left [ \frac{2}{3} (a_1 +a_2) e^{i\delta} -
\frac{1}{3} (2a_1 -a_2) - \left (a^u_4 + a^u_6 \xi^{~}_\pi \right ) \right ]
\right . \nonumber \\
&  & \left . ~~~~ - V^*_{cb} V_{cd} \left [ \sqrt{2} ~ a_1 \kappa + 
\left ( a^c_4 + a^c_6 \xi^{~}_\pi \right ) \right ] \right \} 
T_\pi e^{i\delta_0} \; ,
%             (10)
\end{eqnarray}
where $\delta = \delta_2 - \delta_0$ and $\kappa = S^{\pi D}_0 T_D/T_\pi$.
In obtaining Eq. (10), we have neglected the small quantities of
${\cal O}(|a^{u,c}_4|\cdot |\kappa|)$ and ${\cal O}(|a^{u,c}_6|\cdot |\kappa|)$.
Indeed we find $T_D/T_\pi \approx 3$ for $f_D = 200$ MeV and 
$F^{BD}_0(0) = 0.7$ \cite{Cheng}, with the help of Eq. (5).
Therefore $|\kappa| \sim {\cal O}(|S^{\pi D}_0|) \sim {\cal O}(10^{-2})$
holds. We observe that $A(B^0_d\rightarrow \pi^+\pi^-)_{\bf 0}$ and
$A(B^0_d\rightarrow \pi^0\pi^0)_{\bf 0}$ in 
Eq. (6) can be reproduced from Eq. (10) by
setting $\kappa =0$ and $\delta = \delta_0 =0$. The free parameters
$\kappa$ and $\delta$ are likely to affect the branching ratios of
$B_d \rightarrow \pi\pi$ transitions significantly. For illustration we
take $\kappa = -0.05, -0.05i, 0, +0.05i, +0.05$, respectively, to compute
${\cal B}(B_d\rightarrow \pi\pi)$ as a function of 
$\delta \in [0, 2\pi]$ by using Eq. (7). The numerical
results are shown in Figs. 1 and 2.
%%%%%%%%%%%%%%%%%%%% Fig. 2 %%%%%%%%%%%%%%%%
\begin{figure}[t]
\epsfig{file=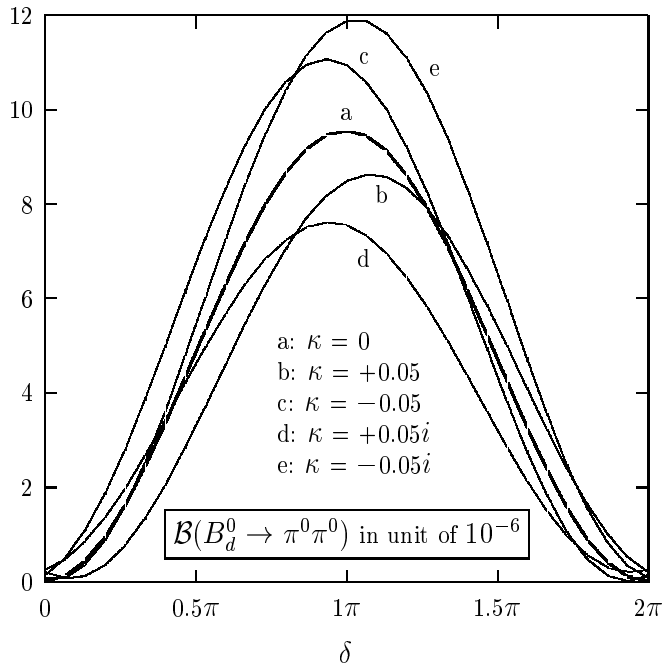,bbllx=1cm,bblly=10cm,bburx=18cm,bbury=27.5cm,%
width=15.5cm,height=17cm,angle=0,clip=}
\vspace{-7.8cm}
\caption{The branching ratio of $B^0_d\rightarrow \pi^0\pi^0$
predicted in the factorization approximation with
different input values of the inelastic rescattering parameter
$\kappa$ and the isospin phase shift $\delta$.}
\end{figure}
%%%%%%%%%%%%%%%%%%%%%%%%%%%%%%%%%%%%%%%%%%%%

Fig. 1 indicates that in the absence of inelastic $\pi\pi \rightleftharpoons
D\bar{D}$ rescattering a good agreement between the theoretical value 
of ${\cal B}(B^0_d\rightarrow \pi^+\pi^-)$ and the CLEO data
invokes $\delta \sim 0.5\pi$ or $1.5\pi$; i.e., there may exist
significant elastic $\pi\pi \rightleftharpoons \pi\pi$ rescattering.
This result is apparently consistent with the analyses made in
Refs. \cite{Hou,Wu}. Taken the final-state $\pi\pi \rightleftharpoons
D\bar{D}$ rescattering effect into account, the naive value of
${\cal B}(B^0_d\rightarrow \pi^+\pi^-)_{\bf 0}$ given before
can be lowered even in the case $\delta =0$: e.g., we obtain
${\cal B}(B^0_d\rightarrow \pi^+\pi^-)/
{\cal B}(B^0_d\rightarrow \pi^+\pi^-)_{\bf 0} \approx 0.7$ by taking
${\rm Re}\kappa =0$ and ${\rm Im}\kappa = 0.05$.
We therefore conclude that both kinds of final-state interactions
are important and non-negligible. Indeed there is a rather large
$(\kappa, \delta)$-parameter space, as shown in Fig. 1, in which
the present data of CLEO, BELLE or BABAR on $B^0_d\rightarrow \pi^+\pi^-$ 
and $\bar{B}^0_d\rightarrow \pi^+\pi^-$ decays can well be accommodated.

Fig. 2 shows our prediction for 
${\cal B}(B^0_d\rightarrow \pi^0\pi^0)$ in the presence of both
$\pi\pi \rightleftharpoons \pi\pi$ and $\pi\pi \rightleftharpoons
D\bar{D}$ rescattering effects. We observe that the naive value of
${\cal B}(B^0_d\rightarrow \pi^0\pi^0)_{\bf 0}$ can be enhanced
up to two orders of magnitude. In the region of $\delta \sim 0.5\pi$
or $1.5\pi$, ${\cal B}(B^0_d\rightarrow \pi^0\pi^0)
\sim 2\times 10^{-6}$ to $8\times 10^{-6}$ is expected for
different values of $\kappa$. Considering the CLEO upper limit 
${\cal B}(B^0_d\rightarrow \pi^0\pi^0) < 5.7 \times 10^{-6}$ \cite{CLEO},
however, we see that the region $\delta <0.5\pi$ or $\delta >1.5\pi$
is more favored
%%%%%%%%%%%%%%%%%%%%%%%%%
\footnote{Note that $\delta \approx 11^\circ$ is expected in the Regge 
model \cite{Gerard}.}.
%%%%%%%%%%%%%%%%%%%%%%%%%
An experimental determination of the
branching ratios of $B^0_d\rightarrow \pi^0\pi^0$ and
$\bar{B}^0_d\rightarrow \pi^0\pi^0$ transitions at the level of $10^{-6}$
should signify significant final-state interactions.
%%%%%%%%%%%%%%%%%%%% Fig. 3 %%%%%%%%%%%%%%%%
\begin{figure}[t]
\epsfig{file=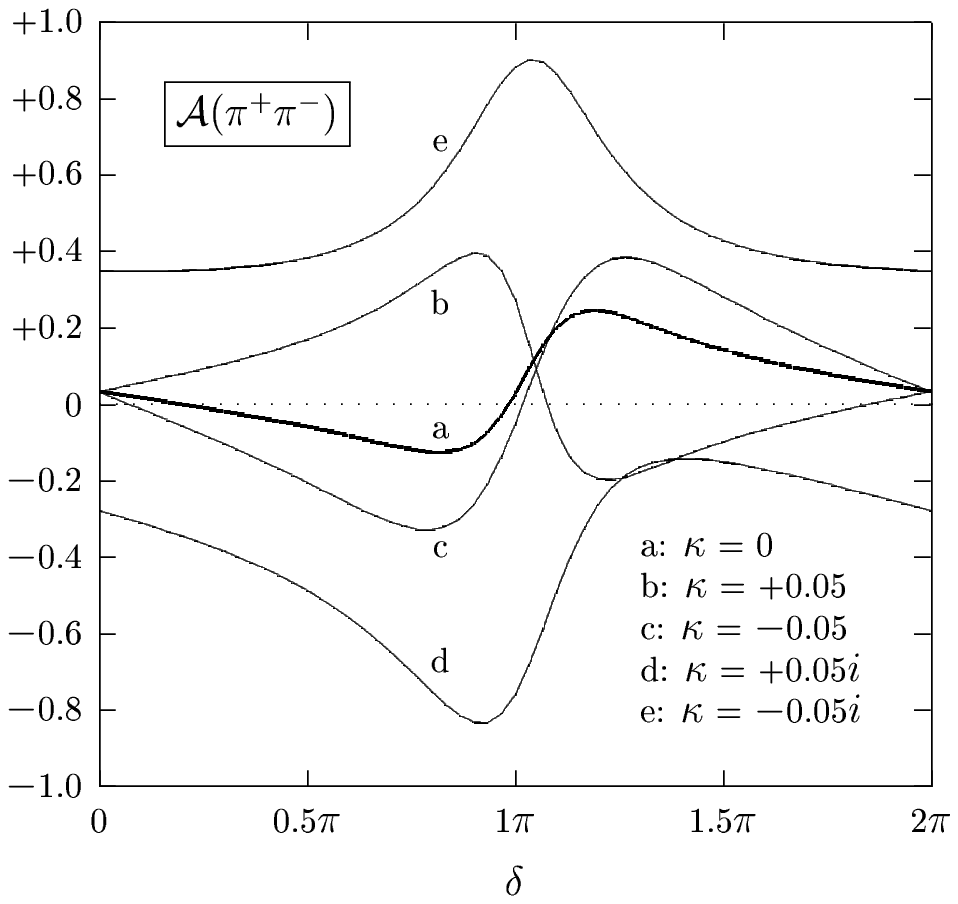,bbllx=1cm,bblly=10cm,bburx=18cm,bbury=27.5cm,%
width=15.5cm,height=17cm,angle=0,clip=}
\vspace{-7.8cm}
\caption{The CP-violating asymmetry of $B^0_d$ vs 
$\bar{B}^0_d\rightarrow \pi^+\pi^-$ decays
predicted in the factorization approximation with
different input values of the inelastic rescattering parameter
$\kappa$ and the isospin phase shift $\delta$.}
\end{figure}
%%%%%%%%%%%%%%%%%%%%%%%%%%%%%%%%%%%%%%%%%%%%
%%%%%%%%%%%%%%%%%%%% Fig. 4 %%%%%%%%%%%%%%%%
\begin{figure}[t]
\epsfig{file=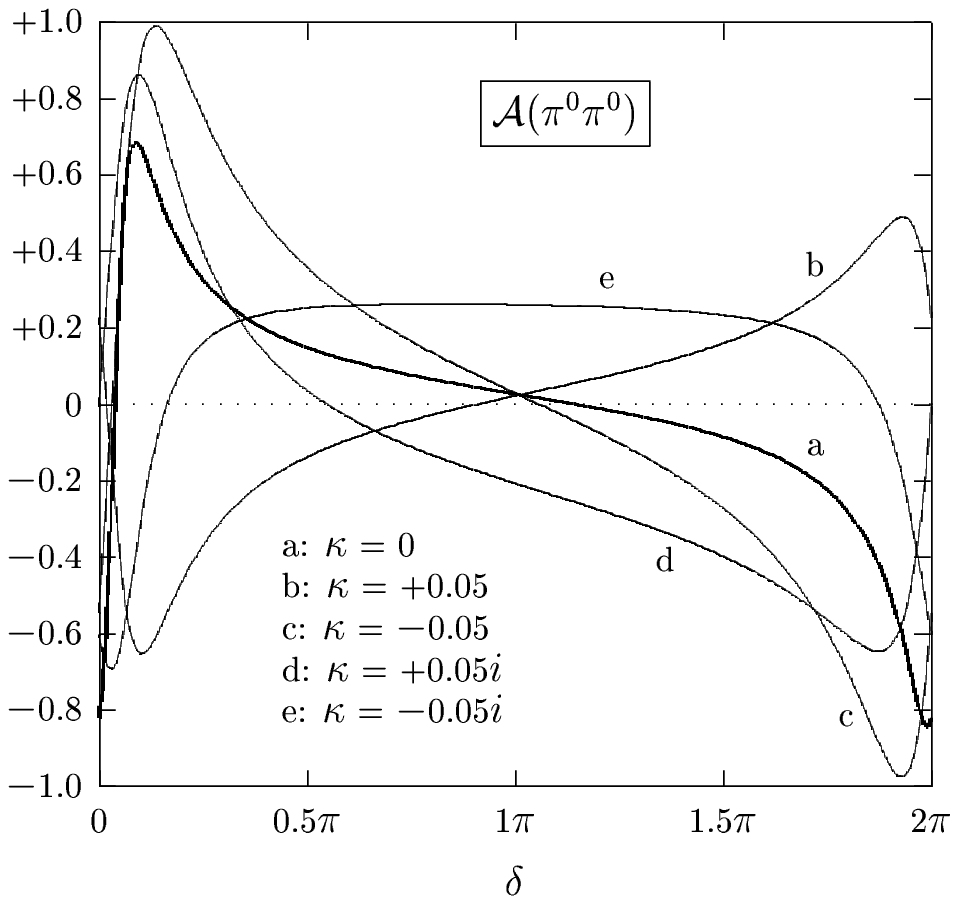,bbllx=1cm,bblly=10cm,bburx=18cm,bbury=27.5cm,%
width=15.5cm,height=17cm,angle=0,clip=}
\vspace{-7.8cm}
\caption{The CP-violating asymmetry of $B^0_d$ vs 
$\bar{B}^0_d\rightarrow \pi^0\pi^0$ decays
predicted in the factorization approximation with
different input values of the inelastic rescattering parameter
$\kappa$ and the isospin phase shift $\delta$.}
\end{figure}
%%%%%%%%%%%%%%%%%%%%%%%%%%%%%%%%%%%%%%%%%%%%

We proceed to calculate the direct CP-violating asymmetries in
$B^0_d$ vs $\bar{B}^0_d\rightarrow \pi^+\pi^-$ and $\pi^0\pi^0$ 
decays, defined respectively as
\begin{eqnarray}
{\cal A}(\pi^+\pi^-) & = & \frac{|A(B^0_d\rightarrow \pi^+\pi^-)|^2
- |A(\bar{B}^0_d\rightarrow \pi^+\pi^-)|^2}
{|A(B^0_d\rightarrow \pi^+\pi^-)|^2 + |A(\bar{B}^0_d\rightarrow \pi^+\pi^-)|^2} \; ,
\nonumber \\ \nonumber \\
{\cal A}(\pi^0\pi^0) & = & \frac{|A(B^0_d\rightarrow \pi^0\pi^0)|^2
- |A(\bar{B}^0_d\rightarrow \pi^0\pi^0)|^2}
{|A(B^0_d\rightarrow \pi^0\pi^0)|^2 + |A(\bar{B}^0_d\rightarrow \pi^0\pi^0)|^2} \; .
%              (11)
\end{eqnarray}
These asymmetries can be observed time-independently on the 
$\Upsilon (4S)$ resonance with a trivial dilution factor
$1/(1+x^2_d) \approx 0.66$ due to $B^0_d$-$\bar{B}^0_d$ mixing \cite{Gronau},
or time-dependently at asymmetric $B$-meson factories running around
the $\Upsilon (4S)$ energy threshold \cite{BF}. The numerical results of
${\cal A}(\pi^+\pi^-)$ and ${\cal A}(\pi^0\pi^0)$ are shown in
Figs. 3 and 4, where we have used the same values as before for the
relevant input parameters. Some comments are in order:

(a) In the absence of final-state interactions at the hadron level
(i.e., $\delta =0$ and $\kappa =0$), the CP asymmetries 
${\cal A}(\pi^+\pi^-) \approx 3\%$ and ${\cal A}(\pi^0\pi^0) \approx -82\%$
are a consequence of the interference between tree-level and
penguin amplitudes, where the non-trivial strong phase shift 
arises from the penguin quark-loop function \cite{Soni}.
Note that we have neglected possible effects from the electroweak
penguins \cite{He}, the space-like penguins \cite{Du95}, 
and the self-interference of different penguin loops, as they are 
generally expected to be insignificant in the transitions under consideration.

(b) If only the $\pi\pi\rightleftharpoons \pi\pi$ rescattering effect
is ``switched on'', ${\cal A}(\pi^+\pi^-)$ undergoes an oscillation
with increasing values of $\delta$ and its magnitude can be as large
as $25\%$ for $\delta \approx 1.2\pi$, while the magnitude of
${\cal A}(\pi^0\pi^0)$ always decreases when $\delta$ deviates from
$0$ or $2\pi$. In this case CP violation remains resulting from
the interference between tree-level and penguin amplitudes, but the
relevant isospin phase differences may play a more important role
than the strong phase shifts induced by penguin loops at the quark level.

(c) If only the $\pi\pi\rightleftharpoons D\bar{D}$ rescattering effect
is ``switched on'', the magnitude of ${\cal A}(\pi^+\pi^-)$ can
remarkably be enhanced (e.g., ${\cal A}(\pi^+\pi^-) \approx 35\%$
for $\kappa = -0.05i$), but that of ${\cal A}(\pi^0\pi^0)$ becomes
smaller than in the case $\kappa =0$. There are two sources of CP
violation: one is the interference between tree-level and penguin
amplitudes, and the other is the interference between two different
tree-level amplitudes as a result of $\pi\pi\rightleftharpoons D\bar{D}$
rescattering. The latter is measured by $\kappa$, whose magnitude and
phase can both affect the CP-violating asymmetries in a significant way.

(d) In general $\pi\pi\rightleftharpoons \pi\pi$ and
$\pi\pi\rightleftharpoons D\bar{D}$ rescattering effects should both be
taken into account. It is then possible to have 
$|{\cal A}(\pi^+\pi^-)| \sim {\cal O}(1)$ and
$|{\cal A}(\pi^0\pi^0)| \sim {\cal O}(1)$ for appropriate values of
the input parameters. While large CP asymmetries are likely to be
measured in $B_d\rightarrow \pi\pi$ decays, to pin down the
true mechanism of CP violation will be very difficult. Furthermore,
the indirect CP-violating signals in such decay modes 
(arising from the interplay of direct decay and $B^0_d$-$\bar{B}^0_d$
mixing) are unavoidably contaminated by significant final-state rescattering 
effects \cite{Xing95,Xing98}. It is therefore a big challenge to extract any
information on the weak CP-violating phases from $B_d\rightarrow \pi\pi$
transitions.

In summary, we have demonstrated the significance of elastic 
and inelastic final-state rescattering effects in $B_d\rightarrow \pi\pi$
decays. Our treatment of the complicated inelastic 
final-state interactions is soley to take the simple 
$\pi\pi \rightleftharpoons D\bar{D}$ rescattering into account, hence
it remains quite preliminary and can only serve for illustration. 
Nevertheless, the present experimental data on $B\rightarrow \pi\pi$
decays can well be understood in our approach without fine-tuning of
the input parameters, and large CP-violating
asymmetries are expected to manifest themselves in such charmless
rare processes. We remark that further effort is desirable towards a
deeper understanding of the dynamics of nonleptonic $B$ decays.

\vspace{0.3cm}
I would like to thank 
X. Calmet, J.N. Ng, X.Y. Pham, and in particular D.S. Yang
for useful discussions. I am also grateful to J.M. G$\rm\acute{e}$rard
and W.S. Hou for useful comments.

\end{document}